\documentclass[aps,prl,amsmath,twocolumn, mssymb,unsortedaddress,superscriptaddress]{revtex4-1}
\usepackage{graphicx}
\usepackage{color}
\usepackage{float}
\usepackage{amsmath}

\newcommand{\beq}{\begin{equation}}
\newcommand{\eeq}{\end{equation}}
\newcommand{\bea}{\begin{eqnarray}}
\newcommand{\eea}{\end{eqnarray}}

\newcommand{\qq}{{\bf q}}

\newcommand{\rr}{{\bf r}}

\newcommand{\la}{{\langle\kern-2.0pt\langle}}
\newcommand{\ra}{{\rangle\kern-2.0pt\rangle}}
\newcommand{\vt}{{\vert\kern-1.0pt\vert}}

\newcommand{\ket}[1]{\vert{#1}\rangle}

\newcommand{\matel}[3]{\langle#1\vert{#2}\vert{#3}\rangle}

\begin{document}

\title{Theory of momentum-resolved phonon spectroscopy in the electron microscope}

\author{R.J. Nicholls}
\email{rebecca.nicholls@materials.ox.ac.uk}
\affiliation{Department of Materials, University of Oxford, Parks Road, Oxford, OX1 3PH, United Kingdom}
\author{F.S. Hage}
\affiliation{SuperSTEM Laboratory, SciTech Daresbury Campus, Keckwick Lane, Warrington, WA4 4AD, United Kingdom}
\author{D.G. McCulloch}
\affiliation{Physics, School of Sciences, RMIT University, Melbourne, VIC 3001, Australia}
\author{Q.M. Ramasse}
\affiliation{SuperSTEM Laboratory, SciTech Daresbury Campus, Keckwick Lane, Warrington, WA4 4AD, United Kingdom}
\affiliation{School of Physics and School of Chemical and Process Engineering, University of Leeds, Leeds LS2 9JT, United Kingdom}
\author{K. Refson}
\affiliation{Department of Physics, Royal Holloway, University of London, Egham TW20 0EX, United Kingdom}
\affiliation{ISIS Spallation Neutron Source, RAL, Chilton, Didcot OX11 0QX, United Kingdom}
\author{J.R. Yates}
\affiliation{Department of Materials, University of Oxford, Parks Road, Oxford, OX1 3PH, United Kingdom}


\begin{abstract}
We provide a theoretical framework for the prediction and interpretation of momentum dependent phonon spectra due to coherent inelastic scattering of electrons. We complete the approach with first principles lattice dynamics using periodic density functional theory and compare to recent electron energy loss measurements on cubic and hexagonal boron nitride performed within a scanning transmission electron microscope (STEM). The combination of theory and experiment provides the ability to interpret momentum dependent phonon spectra obtained at nanometer spatial resolution in the electron microscope.
\end{abstract}


\maketitle


\section{Introduction}

The quantitative description of the thermal physics of solid materials in terms of quantized lattice vibrations, phonons, is one of the major achievements of condensed-matter physics in the 20th century.  Lattice dynamics is central to the theories of phenomena including structural phase transitions, superconductivity, thermal expansion, thermal conductivity, stability of polymorphs and much more.  Laboratory techniques to measure phonon spectra using light including infra-red and Raman spectroscopy are powerful and widely deployed across laboratories, but the energy-momentum relation of the photon probe restricts the interaction with phonons to involve essentially zero momentum transfer. Consequently only a subset of phonon modes at the long-wavelength limit may be measured using optical probes.

Inelastic neutron scattering (INS), pioneered by Bertram Brockhouse~\cite{brockhouse95}, was the major development which enabled the full measurement of phonon spectra at all phonon wavevectors - the first momentum-resolved spectroscopy. This was followed by inelastic X-Ray scattering (IXS)~\cite[and citations therein]{Baron2009,Krisch2017}. Such techniques have been the mainstay of phonon spectroscopy in crystalline solids for half a century. However their application is limited by the scarcity and expense of INS and IXS spectrometers, which must be based at reactor, accelerator or synchrotron sources. The requirement to grow single crystal specimens also limits their widespread use, particularly in the case of neutrons where crystal sizes of 20-1000$mm^3$ are needed. The spatial resolution of INS is larger than $1 cm$, and while X-Ray spot sizes in the micrometer range can easily be obtained, counting rates and experimental timescales mostly preclude spatially-resolved studies.

Electrons have been used since the 1960s in a reflection geometry to measure the average surface response of materials~\cite{propst,ibach}.  Recent advances in source monochromation mean that it is now possible to measure phonon spectra in a transmission electron microscope using electron energy-loss spectroscopy (EELS) with a resolution of $15 meV$ or better~\cite{krivanek}.  This adds a complementary technique to the methods above, with the additional advantages of nanometer spatial resolution~\cite{krivanek,dwyer} of the phonon spectrum, alongside atomically resolved chemical and structural analysis, all within the same instrument.

The geometrical constraints in reflection EELS mean that the theoretical treatments used are not applicable to transmission EELS.  For the transmission geometry, two different scattering regimes have been identified by Dwyer $et~al.$~\cite{dwyer}: dipole scattering and localised vibrational scattering.  Dipole scattering involves small momentum transfer and Radtke $et~al.$~\cite{radtke} have used density functional theory to model EEL spectra in this regime.  To interpret momentum dependent spectra, we must consider the localised vibrational scattering regime.  

The theory of INS from phonons was developed in a very general formalism by Leon van Hove~\cite{vanhove}, and can be adapted to any radiative probe for which the interaction Hamiltonian is known. In this paper we present its extension to coherent inelastic scattering of electrons from phonons and apply it to the case of momentum-resolved EELS experiments performed in a scanning transmission electron microscope (STEM).  This formalism enables the prediction of scattering cross section as a function of momentum and energy transfer and makes possible a quantitative comparison with EELS experiments.  It reveals the fundamental physics shared between inelastic scattering of electrons, neutrons and photons, and attempts to unify the theories of EELS, INS and IXS.  In contrast, previous work has looked at specific cases~\cite{rez,garcia,supp}, been used to interpret the $q=0$ modes in a nanocube~\cite{lagos,hohenester}, has looked at spatial the effects of beam geometry~\cite{forbes} or dealt with the dipole ($q\approx0$) scattering regime~\cite{radtke,govyadinov}.  We apply this general method to predict the phonon contribution to the EEL spectrum of two polymorphs of boron nitride and make a direct comparison to their recently-measured momentum-resolved spectra~\cite{hage_1}.

\section{Scattering Factor formalism}

As the energy transfer that occurs in the scattering process is small compared to that of the scattered particle, the double differential cross section is given by the Born approximation as  ~\cite{sturm}
\begin{widetext}
\begin{equation}
\label{ddc_1}
\frac{d^2\sigma}{d\Omega{}dk_1} = \frac{1}{N}\frac{N_0V\sum_{n_0,n_1}P_{n_0}k_1^2|\matel{n_0,\mathbf{k}_0}{H_{inter}}{n_1,\mathbf{k}_1}|^2\delta(E_{n_0}+E_0-E_{n_1}-E_1)}{(2\pi)^2\hbar(j_0)_z}
\end{equation}
\end{widetext}
where $n_0$ and $n_1$ are the initial and final states of the material with energies $E_{n_0}$ and $E_{n_1}$ respectively, $\mathbf{k}_0$ and $\mathbf{k}_1$ are the initial and final states of the scattered particle with energies $E_0$ and $E_1$ respectively, $H_{inter}$ is the Hamiltonian for the interaction of the particle with the material, $(j_0)_z$ is the current density of the beam of particles in the $z$-direction, $P_{n_0}$ is the probability of finding the material in state $n_0$ before scattering, $N$ is the number of scatterers in the material, $N_0$ is the number of electrons in state $k_0$ and $V$ is the volume of the unit cell.  The scattered particles could be photons, neutrons or fast electrons.  For the different particles, the form of the interaction Hamiltonian and expressions for the current density are different.  A fast electron will interact with both the electrons and nuclei in the sample and the interaction Hamiltonian can be written as 
\begin{equation}
\label{ham}
H_{inter}(\mathbf{r}) = \frac{-e}{4\pi\epsilon_0}\int\frac{\rho_{tot}(\rr')}{|\mathbf{r}-\rr'|}d\rr' 
\end{equation}
where $\mathbf{r}$ is the fast electron position, $\rr'$ is the material co-ordinate, $e$ is the magnitude of charge of an electron and $\rho_{tot}$ is the total charge density containing both the nuclear and electronic contributions. Here we have assumed a material with no spin density in the ground state. Retardation effects have also been neglected. Before and after the scattering event, the fast electron and material do not interact, so we can write $\ket{n_0,\mathbf{k}_0} = \ket{n_0}\ket{\mathbf{k}_0} $ and $\ket{n_1,\mathbf{k}_1} = \ket{n_1}\ket{\mathbf{k}_1} $.  By including the full interaction Hamiltonian in Eq.~\ref{ddc_1}, defining the momentum transfer from the fast electron to the sample as $\mathbf{q} = \mathbf{k}_0-\mathbf{k}_1$, approximating the fast electron as a plane wave, writing the energy transfer as $\hbar\omega$ and writing $(j_0)_z$ as $\frac{N_0}{V}\frac{\hbar{}k_0}{m}$, Eq.~\ref{ddc_1} becomes
\begin{widetext}
\begin{equation}
\label{ddc_2}
\frac{d^2\sigma}{d\Omega{}dE_1}  =  \frac{m^2e^2}{\hbar^4q^4\epsilon_0^24\pi^2}\frac{k_1}{k_0}\frac{1}{N}\sum_{n_0,n_1}P_{n_0}|\matel{n_0}{\int{}d\rr'e^{-2\pi{}i\mathbf{q}.\rr'}\rho{}_{tot}(\rr')}{n_1}|^2\delta(E_{n_0}-E_{n_1}+\hbar\omega) 
\end{equation}  
\end{widetext}
Following Van Hove~\cite{vanhove}, the double differential cross section can be written in terms of a scattering function, $S(q,\omega)$  
\begin{equation}
\label{ddc_3}
\frac{d^2\sigma}{d\Omega{}dE_1} = \frac{m^2e^2}{\hbar^4q^4\epsilon_0^24\pi^2}\frac{k_1}{k_0}\frac{1}{N}S(q,\omega)
\end{equation}

To determine $S(q,\omega)$ we follow the general approach of Sinha~\cite{sinha} and Burkel~\cite{burkel} who considered X-Ray scattering from phonon vibrations. We first assume that the total charge density can be expressed as a sum of atomic charge densities. This is clearly a major simplification, and we return to this approximation later. We also assume harmonic lattice dynamics in a crystal and can hence express the lattice vibrations expressed as phonon eigenvectors.  By neglecting processes involving multiple phonons, we obtain an expression for $S(q,\omega)$ for the creation of phonons by a fast electron:
\begin{widetext}
\begin{equation}
\label{sqw_2}
S(q,\omega) = \Big{|}\sum_iF(\qq,Z_i) e^{-W_i(\mathbf{q})}[\mathbf{q}.\mathbf{e}_i(\mathbf{q}_0,j)]M_i^{-1/2}e^{i\mathbf{q}.\mathbf{r}_i}\Big{|}^2\frac{1}{\omega_{\mathbf{q}_0j}}\delta(\omega-\omega_{\mathbf{q}_0j})
\end{equation} 
\end{widetext}
where there are $i$ atoms per unit cell at positions $\mathbf{r}_i$, $M_i$ and $Z_i$ are the mass and atomic number of atom $i$, $\mathbf{e}_i(\mathbf{q}_0,j)$ is the phonon eigenvector with wavevector $\mathbf{q}_0$ (defined in the first Brillouin Zone) and polarisation branch $j$ at atom $i$ and $e^{-2W_i(\mathbf{q})}$ is the Debye-Waller factor. $F(\qq,Z_i)$ is given by
\begin{equation}\label{eq:bigF}
F(\qq,Z_i)=f_{atom,i}(\mathbf{q})+Z_ie
\end{equation}
where $f_{atom,i}(\mathbf{q})$ is the atomic form factor.  

Eq.~\ref{sqw_2} contains a term, $\mathbf{q}.\mathbf{e}_i(\mathbf{q}_0,j)$, which is the dot product between the momentum transfer and the phonon eigenvector.  This means that only modes with motion in the same direction as the momentum transfer will appear in the spectra.  Rez~\cite{rez} obtained an expression for the differential cross-section in the case of a fast electron interacting with a stretch vibration in a diatomic molecule and there are clear similarities between his equation for that specific case and our general equation.  Rez commented on the implications of the $\mathbf{q}.\mathbf{e}(\mathbf{q}_0,j)$ term and also pointed out the connection between the cross section and the loss function (the imaginary part of $-1/\epsilon{}(E,\mathbf{q})$ where $\epsilon{}(E,\mathbf{q})$ is the energy and wavevector-dependent dielectric function); more details of the relationship can be found in Ref.~\cite{sturm}.

The scattering function formalism developed above enables us to go beyond simply comparing momentum-dependent EEL spectra to phonon bandstructures and understand the relative contributions of the different modes to the spectra.  It has been developed to understand scattering in which there is finite momentum transfer, a regime known as the localised vibrational scattering regime~\cite{dwyer}; a correct treatment of the $q=0$ term should also include dipole scattering~\cite{dwyer,radtke,govyadinov} rather than solely impact scattering.

\subsection{Comparison with the scattering factor formalism for X-Rays and neutrons}

The single-phonon scattering factor obtained for the interaction of electrons with phonons is very similar to those obtained for X-Rays and neutrons, highlighting the complementary nature of the techniques. 
For X-Rays, the double differential cross section for energy loss can be written as~\cite{burkel,sinha}

\begin{equation}
\frac{d^2\sigma}{d\Omega{}dE_1} = \frac{e^4}{m^2c^4}\frac{k_1}{k_0}|\mathbf{\epsilon}^*_0.\mathbf{\epsilon}_1|^2S(\mathbf{q},\omega)
\end{equation}

where $\mathbf{\epsilon}_0$ and $\mathbf{\epsilon}_1$ are the polarisation vectors of the incoming and outgoing photons.  The scattering function, $S(\mathbf{q},\omega)$ is given by 

\begin{widetext}
\begin{equation}
S(\mathbf{q},\omega) = \Big{|}\sum_if
_{atom,i}(\mathbf{q})e^{-W_i(\mathbf{q})}[\mathbf{q}.\mathbf{e}_i(\mathbf{q}_0,j)]M_i^{-1/2}e^{i\mathbf{q}.\mathbf{r}}\Big{|}^2\frac{<n_{\mathbf{q}_0j}+1>}{\omega_{\mathbf{q}_0j}}\delta(\omega-\omega_{\mathbf{q}_0j})
\end{equation}
\end{widetext}

To obtain the neutron scattering function, the atomic form factor is replaced by the Fermi scattering length, $b$~\cite{burkel,squires}.  In all three cases, X-Rays and neutrons and electrons, there is a $\mathbf{q}.\mathbf{e}_i(\mathbf{q}_0,j)$ term, showing that the same bands contribute to the spectra produced by the scattering of fast electrons, X-Rays and neutrons.  

The double differential cross section for neutrons varies as $q^2$.  For X-Rays the $q^2$ dependence is counteracted by the atomic form factor, which also has a $q$ dependence.  Consequently there is an optimal range of $q$ vectors for which the cross-section will be greatest.  The range will be material dependent but is generally well outside the first Brillouin Zone.  Eq.~\ref{sqw_2} appears to have a $q^{2}$ dependence which, when combined with the $q^{-4}$ term in Eq.~\ref{ddc_3}, results in a $q^{-2}$ dependence.  However, $F(\mathbf{q},Zi)$ is also q-dependent and the overall variation depends on the degree of ionicity in the crystal~\cite{rez}.  In addition, for the $q=0$ case to be described correctly, an additional dipole term should be included.  In contrast, the cross section for neutrons has a $q^2$ dependence and for X-Rays there is an optimal range of $q$-values for data collection.  From a practical point of view, this means that experimental data using these techniques is rarely collected from the first Brillouin Zone as large values of $q$ will give a greater signal.  When using electrons, the signal will be strongest within the first Brillouin Zone.  

As mentioned in the Introduction, the spatial resolution achievable with fast electrons is much greater than that of either X-Rays or neutrons.  The energy and momentum resolution, however, are not as good.  In the experiments carried out here, the energy resolution was 18-40$meV$ (increasing with increasing q) and the momentum resolution was $\pm{}0.5 \AA^{-1}$.  This compares with typical energy and momentum resolutions of 0.6-6$meV$ and $0.01-0.1 \AA^{-1}$ for X-Rays~\cite{Baron2009,esrf}.  For neutrons, the energy and momentum resolution depends on the application and specific instrument and it can be defined using a 3D resolution ellipsoid~\cite{cussen}.  Some reactor-based sources can achieve $\mu{}eV$ resolution~\cite{neutron}.  The high-resolution instrument MAPS~\cite{isis} can achieve an energy resolution of 0.4$meV$ for neutrons with incident energies of 25$meV$, but this increases to 30$meV$ for incident energies of 2000$meV$.  Infra-red and Raman spectroscopies are both powerful and widely-used methods for measuring phonons but, rather than measuring phonon dispersions, they are essentially limited to probing $q=0$ and so provide complementary ways of exploring $q=0$ transitions.

\section{Application to cubic and hexagonal Boron Nitride}

\subsection{Experimental details}
Based on earlier experimental procedures for acquiring momentum resolved core and valance EEL spectra in the (scanning) transmission electron microscope ((S)TEM) (see e.g. Refs.~\cite{leapman,hage} and references therein), we use the electromagnetic scan coils of the STEM to accurately control the range, magnitude and direction of momentum transfers accepted by the circular spectrometer entrance aperture.  Thus our momentum resolved spectra are acquired serially~\cite{hage}, rather than in parallel~\cite{leapman}.  By carefully balancing the simultaneously achievable spatial and momentum resolutions, we probed phonon dispersions along high symmetry directions of the first Brillouin zones of cubic and hexagonal BN, using a $\sim 1nm$ electron probe, a momentum resolution $\Delta{}q = \pm{}0.5 \AA^{-1}$ and an energy resolution $\Delta{}E =$ 18-40$meV$ (increasing with increasing q)~\cite{hage_1}.  

All experimental work was carried out on a Nion UltraSTEM100MC dedicated STEM~\cite{krivanek2009}, equipped with a Gatan Enfinium EEL ERS spectrometer (optimised with high stability electronics). The microscope was operated at an acceleration voltage of $60kV$, in order to minimise electron beam induced irradiation damage. No post-acquisition de-noising or deconvolution routines were used for any experimental spectrum. In each spectrum shown in Fig.~\ref{spec} (and in the Figures in the Supplemental material), the zero loss peak (ZLP) tail contribution was subtracted by fitting a power law function over an energy loss window preceding the lowest energy loss peak.  In addition, the intensity of the experimental spectrum for each $q$ vector has been normalised.  

In inelastic neutron spectroscopy, it is necessary to determine the full resolution ellipsoid, expressing energy and momentum resolutions in terms of energy loss and momentum transfer~\cite{cussen}. In our case, the momentum resolution ($\Delta{}q = \pm0.5 \AA^{-1}$) is limited by beam convergence and spectrometer acceptance angles (both with a half-angle of $3 mrad$), while the energy resolution ($\Delta{}E$) is approximately constant as a function of momentum transfer. Experimentally, $\Delta{}E$ is measured as the full width at half maximum (FWHM) of the quasi elastic ZLP. Due primarily to instrumental instabilities, along with electron-atom Compton scattering~\cite{argentero}, the measured ZLP FWHM effectively increases from $18-20 meV$ (near the optical limit) to $30-40 meV$ for larger momentum transfers. The ZLP broadening due to instrumental instabilities increases with q due to the significantly longer exposure times required at higher q (from $0.1s$ near the $\Gamma$ point up to $90s$ at the Brillouin zone boundaries). In light of the above mentioned factors, the more involved procedure for determining the full resolution ellipsoid~\cite{cussen} was not deemed appropriate. For a more detailed discussion of the experimental procedure and associated parameters, see reference~\cite{hage_1}. 

\subsection{Details of First-Principles calculations}

The phonon eigenvalues and dispersions for cubic boron nitride (cBN) and hexagonal boron nitride (hBN) have been calculated with density functional theory code CASTEP~\cite{castep} using norm-conserving pseudopotentials and the PBE exchange-correlation functional.  A geometry optimisation was first carried out on both crystal structures using a cut-off energy of $850eV$, a k-point sampling of $2\pi{}\times0.03\AA^{-1}$ and the structures were optimised until the forces on each atom did not exceed $0.001eV/\AA$.  Phonon dispersions were then calculated using density functional perturbation theory (DFPT) with Fourier interpolation used to calculate the dynamical matrices on a finer grid~\cite{castep_dfpt}.  The DFPT calculations used a cut-off energy of $850eV$, a k-point spacing of $2\pi{}\times0.03\AA^{-1}$, a phonon k-point spacing of $2\pi\times0.07\AA^{-1}$ and a fine phonon k-point path spacing of at most $2\pi\times0.05\AA^{-1}$.  The numerical parameters were all carefully checked so that their value did not influence the final phonon dispersions.  

\subsection{Interpretation of the experimental spectra using the scattering factor formalism}

\begin{figure*}
\includegraphics[scale=0.5]{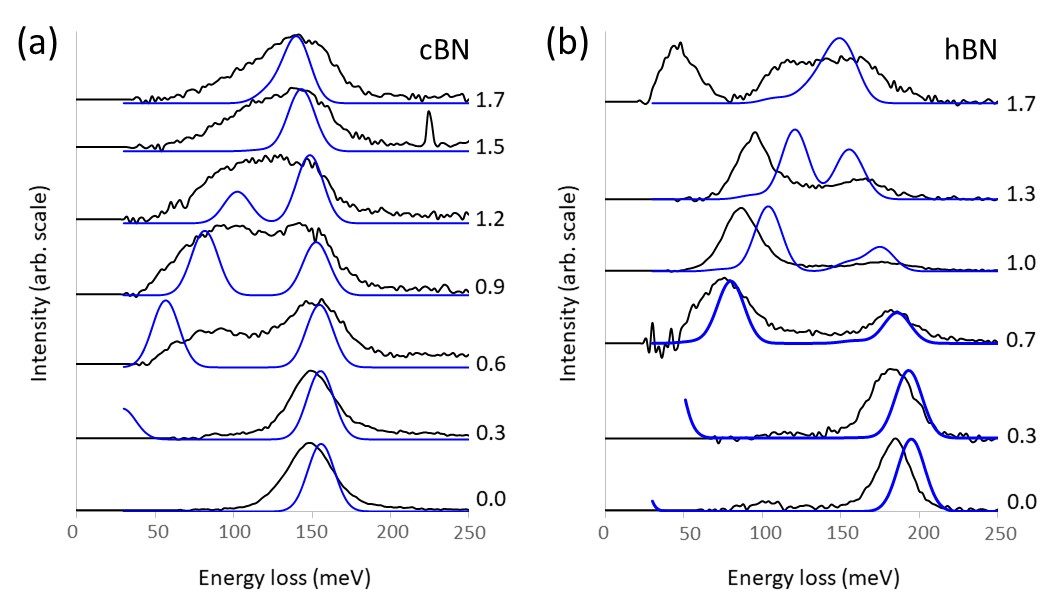}
\caption{Experimental (black) and simulated (blue) phonon EEL spectra for (a) the $\Gamma{}~-$~X direction in cubic BN and (b) the $\Gamma{}~-$~K direction in hexagonal BN.  Each spectrum is labelled by the corresponding $q$ vector in units of $\AA^{-1}$, given to one decimal place.}
\label{spec}
\end{figure*}

Momentum resolved phonon EEL spectra from both cBN and hBN are shown in Fig.~\ref{spec}.  These spectra can be understood using the scattering factor obtained above for the interaction of a fast electron with the vibrational modes of the material.  The effects of charge transfer are significant in BN polymorphs, and as an initial approach to incorporating these effects we replace $F(\qq,Z_i)$ in Eq.~\ref{eq:bigF} with
\begin{equation}
F(\qq,Z_i)=\frac{f_{atom,i}(\mathbf{q})(Z_i-Z^*_i)}{Z_i}+Z_ie
\end{equation}
where $Z^{*}_{i}$ is the Mulliken charge computed for atom $i$. Details of the effect of this approximation on the simulated EEL spectra are given in~\cite{supp}.

The quantity that has been calculated for comparison with experiment is 
\begin{widetext}
\begin{equation}
\label{calc}
\mathcal{J}=\frac{1}{q^4}\Big{|}\sum_i\Big{(}\frac{f_{atom,i}(\mathbf{q})(Z_i-Z^*_i)}{Z_ie}+Z_i\Big{)}e^{-W_i(\mathbf{q})}[\mathbf{q}.\mathbf{e}_i(\mathbf{q}_0,j)]M_i^{-1/2}e^{i\mathbf{q}.\mathbf{r}_i}\Big{|}^2\frac{1}{\omega_{\mathbf{q}_0j}}
\end{equation}
\end{widetext}
where the phonon eigenvectors and frequencies, Debye-Waller factors and Mulliken charges are computed using DFT, and atomic-form factors are taken from the literature. $\mathcal{J}$ can be thought of as a relative intensity and it tells us which of the different modes contribute towards the spectrum and by how much compared to the other modes. The calculated EEL spectrum is constructed by combining Gaussians centred on each of the phonon energies, scaled by $\mathcal{J}$.  The FWHM for the Gaussians was $20 meV$, similar to the experimental resolution.  For modes which have an eigenvector orthogonal to $q$, $\mathcal{J}$ will be zero and they will not contribute to the spectrum.  

\begin{figure*}[t]
\includegraphics[scale=0.2]{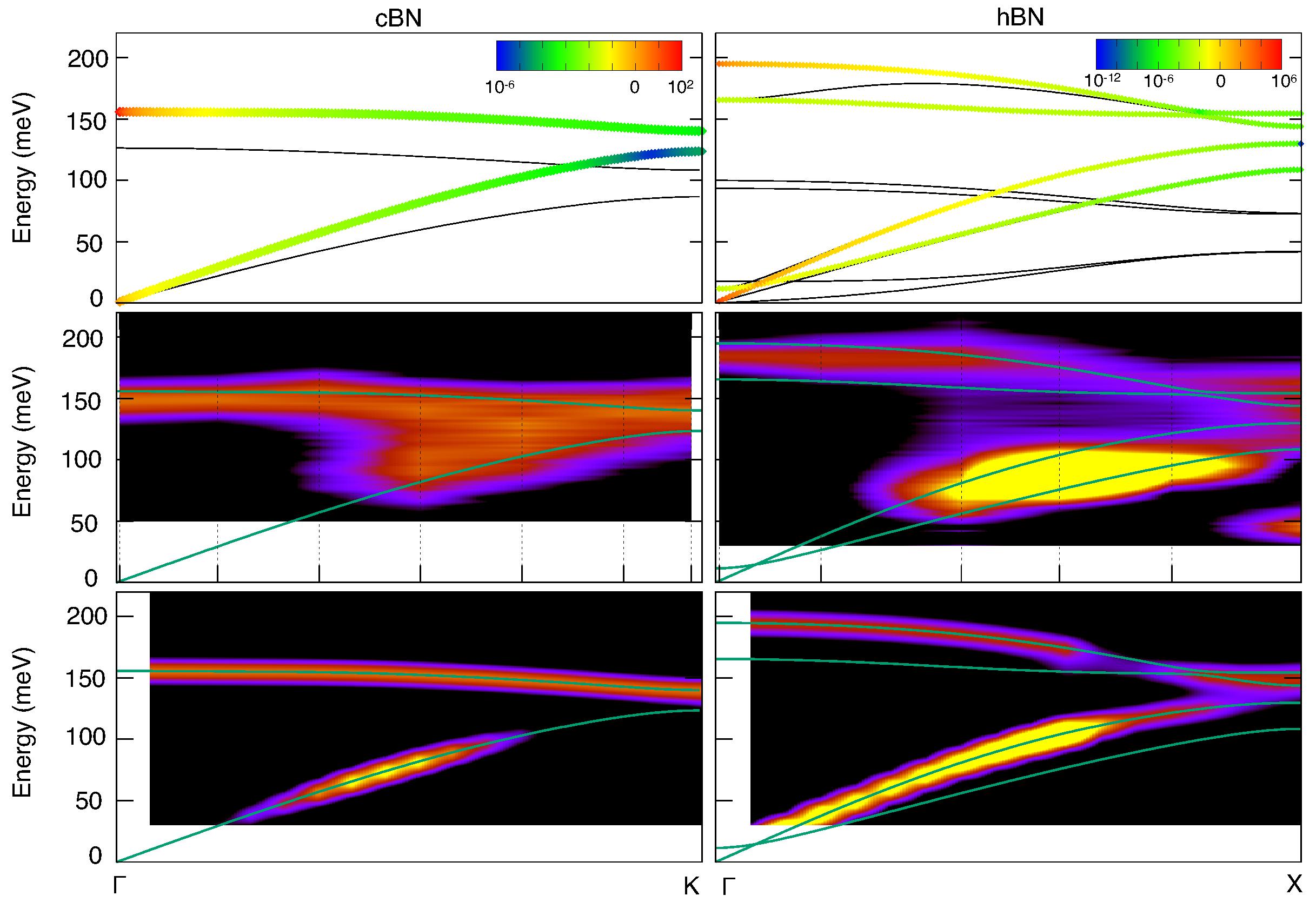}
\caption{Comparison of experimental and computed dispersion relations for cBN (left) and hBN (right). Upper panel: Calculated phonon dispersion spectra. The bands are coloured according to their intensity on a $log_{10}$ scale. Inactive bands are shown in black. Middle Panel: Experimental intensity, normalised by the value of the intensity in the upper branch. The momenta at which the data was recorded is shown by dash vertical lines and the plot is generated by interpolating between the datapoints. The computed DFT band structure is shown in green. Lower panel: Computed intensity, normalised by the value of the intensity in the upper branch. }
\label{spectra1}
\end{figure*}

\begin{figure*}
\includegraphics[scale=0.5]{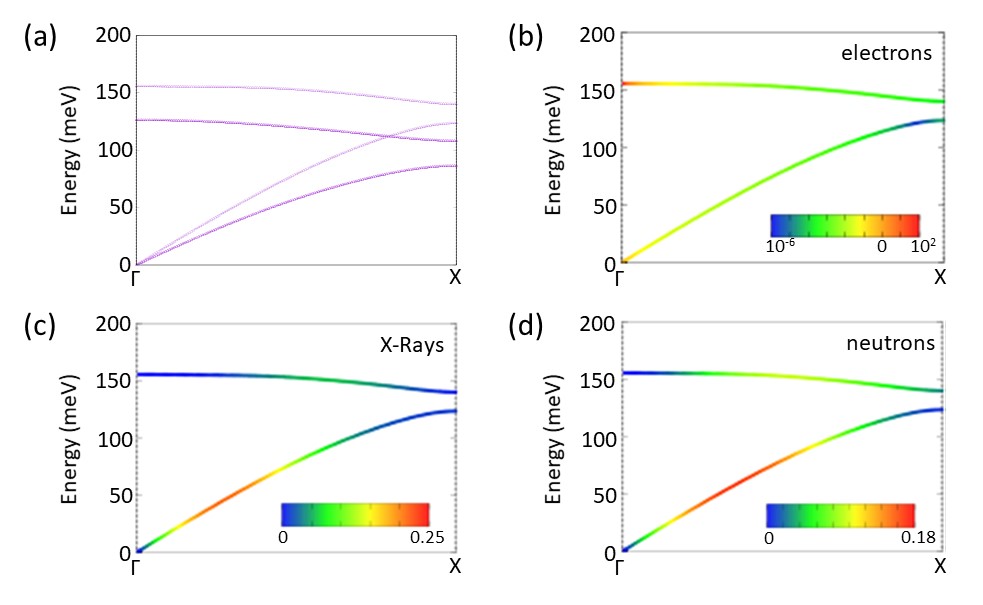}
\caption{Calculated phonon spectra for the $\Gamma{}~-$~X direction in cBN (a) and the relative intensities for the contributing bands in the case of scattering of fast electrons (b), X-Rays (c) and neutrons (d).  Note: in the case of scattering of electrons the contributions of the bands is shown on a $log_{10}$ scale}
\label{cBN_particles}
\end{figure*}

Simulated spectra for cBN are included in Fig.~\ref{spec}a and a further comparison between spectra in the $\Gamma{}~-$~K direction is included in~\cite{supp}.  For comparison with experiment, the simulated loss function has been scaled to match the maximum in the experimental data.  As the $q=0$ term is not well defined, a spectrum has been simulated for $q=0.01$ for comparison with the experimental data.  The $q=0$ experimental spectrum will have a dipole term, which has not been included here, as well as contributions from small values of momentum transfer as a result of the experimental geometry .  Fig.~\ref{spectra1} shows the corresponding part of the phonon dispersion with the colour of the modes corresponding to how much the modes contribute to the spectrum.  Due to the large variation in intensity across the Brillouin zone, it has been plotted on a $log_{10}$ scale.  As can be seen from the figure, only two of the six phonon modes predicted by DFT for cBN contribute to the spectra, one of these is an optical branch and the other an acoustic branch.  For the other modes, the atomic motion is perpendicular to $\mathbf{q}$.  Fig.~\ref{cBN_particles} shows the different relative contributions of the two bands for the different scattering particles (electrons, X-Rays and neutrons).  The variation in intensity is much greater in the case of fast electron scattering.  Four modes contribute in the $\Gamma{}~-$~K direction~\cite{supp}.  In that case, a lower energy mode dominates at higher values of $q$ and the spectrum appears to shift as $|\mathbf{q}|$ increases.     

There is some discrepancy between the experimental and simulated spectra in Fig.~\ref{spec}.  There are two approximations in the simulations which are likely to account for this.  The first is that the simulations have been carried out for a single $q$-value whereas the momentum resolution of our experimental data is $\pm{}0.5 \AA^{-1}$.  The second is the simple model that has been used for charge transfer.

Experimental and simulated data for the $\Gamma~-$~K direction in hBN are shown in Fig.~\ref{spec}b.  The corresponding part of the Brillouin zone showing the contribution of the different modes is shown in Fig.~\ref{spectra1}.  In this case, four of the twelve DFT-predicted modes contribute to the spectra.  Previous work by Serrano $et~al.$~\cite{serrano} has shown good agreement between DFT phonon bandstructures and IXS data from hBN as well as with published  reflection EELS data from~\cite{rokuta}.  Our DFT phonon bandstructures are very similar to those reported by Serrano $et~al.$~\cite{serrano} but the agreement between our simulated and experimental data for hBN is not as good in the cBN case and this is likely to be due partly to our treatment of the so-called LO-TO splitting.  In an infinite crystal, the longitudinal optical (LO) mode is blue-shifted by the interaction between macroscopic electric fields generated by displaced ions as $q\rightarrow{}0$ and the ionic charges.   In a crystal of finite thickness, joint electromagnetic cavity and phonon modes known as phonon polaritons appear with energies intermediate between the LO and TO values.  These modes display strong dependence when thickness is comparable to the optical wavelength. Our calculations have included the LO-TO splitting expected for an infinite crystal whilst the experimental data was collected from a crystal of a thickness where the phonon polaritons are expected to show significant thickness dependence~\cite{hage_1,govyadinov}.  Michel and Verberck \cite{michel} calculated the phonon dispersion of hBN multilayers. Their work shows this effect will only affect the upper two branches that contribute towards the spectrum.  Near the $\Gamma$-point the two branches are further apart in the case of an infinite crystal and the difference between a multilayer and 3D crystal decreases as $\mathbf{q}\rightarrow{}\mathbf{K}$.  Theoretically, the upper-most branch dominates near the $\Gamma$-point and so the simulations will over-estimate the peak position.  For the other values of q, the lower branch dominates and so the simulations will be less affected.  This is what we see in Fig.~\ref{spec}b. 

Another factor contributing to the discrepancy between simulation and experiment is the experimental geometry.  The experimental momentum resolution results in data being collected over a small range of $q$ vectors; this is currently not included in our calculations.  In addition our calculations include in-plane contributions only whilst the curvature of the Ewald Sphere will mean that the contribution of modes with an out-of-plane $q$ component will increase as $q$ increases.  This is seen in our data where, for larger values of $q$, the match between experiment and theory is less good.  The experimental peak positions are close to phonon energies in the dispersion, but not ones that would be expected to contribute towards the spectrum due to the $\mathbf{q}.\mathbf{e}(\mathbf{q}_0,j)$ term.  The finite size of the probe may also have an effect on the spectrum with local inhomogeneities, such as defects, resulting in breaking of symmetry making the $\mathbf{q}.\mathbf{e}(\mathbf{q}_0,j)$ term becomes non-zero.  
Data showing the $\Gamma~-$~M direction shows similar trends (see~\cite{supp}).  

\section{Conclusion}

The scattering function formalism developed here highlights the fundamental similarities between the scattering of electrons, neutrons and X-Ray, as well as the differences resulting from the Coulombic interaction.  In addition to this, the experimental set up used to collect EELS data means that finite momentum and spatial resolution will also possibly need to be considered when interpreting experimental data.  

In this paper we have formulated a general expression for the interaction of a fast electron with phonon vibrations inside a STEM.  We have applied this approach to understand the differences in momentum resolved EEL spectra from different polymorphs of BN.  The simulated spectra match well with the experimental data and allow us to understand which modes are contributing to the spectra.  This is a general approach and will allow interpretation of experimental data from a large variety of materials.  

\section{Acknowledgements}
\begin{acknowledgments}
SuperSTEM is the UK Engineering and Physical Sciences Research Council (EPSRC) National Research Facility for Advanced Electron Microscopy.  RJN gratefully acknowledges financial support from the EPSRC, grant EP/L022907/1.  Raw experimental data were generated at the SuperSTEM Laboratory, with $ab~initio$ modelling carried out at the University of Oxford.  Source data is available from the corresponding authors upon reasonable request.  The authors thank Michael Krisch for his helpful comments on the manuscript.   
\end{acknowledgments}

\bibliography{Phonon_EELS_BN_2}

\end{document}